\documentstyle[twoside,fleqn,mssl_workshop,psfig]{article}

\begin{document}

\title{X-RAY SPECTROSCOPY OF COOL STARS}

\author{M. G\"udel 
\address{Paul Scherrer Institut, W\"urenlingen and Villigen, CH-5232 Villigen PSI, Switzerland}
}

\begin{abstract}
High-resolution X-ray spectroscopy has addressed not only various topics in coronal physics of
stars, but has also uncovered important features relevant for our understanding of stellar evolution
and the stellar environment. I summarize recent progress in coronal X-ray spectroscopy and in particular also
discuss new results from studies of X-rays from pre-main sequence stars.
\end{abstract}

\maketitle

\section{Introduction}

X-rays from cool stars are a manifestation of magnetic fields generated
by the internal dynamos. This by now classical view
has not changed despite the discovery of new X-ray phenomenology and the development
of new theoretical concepts with which to explain the production of X-rays around
stars (e.g., accretion, outflows, fluorescence, etc).

However, with  increasing detector sensitivities and the advent of high-resolution spectroscopy,
``stellar environmental issues'' have become of central interest in 
the area of pre-main sequence stars. The interplay between accretion disks and stellar 
magnetic fields leads to disk instabilities, ionization and heating
of circumstellar material, chemical reactions in cool molecular gas, accretion flows 
from the disk to the star, outflows, and jets,  in all of which X-rays either
play a key role or at least provide important diagnostic access. 

Limitations of present-day instrumentation are, however, evident. The
following sections summarize selected topics of X-rays from cool stars that have been addressed 
in recent years using high-resolution spectroscopy  from {\it XMM-Newton} and 
{\it Chandra}.

\section{Coronal magnetic structure}

To derive location, structure, and extent of stellar magnetic fields,  
only radio interferometric methods are available for direct imaging, with their own 
severe limitations. X-ray photometric studies have made use of rotational modulation or 
of eclipses  (e.g., \cite{ottmann94,guedel95,kuerster97,audard01,marino03}).
While appealing for specific studies of individual objects, such methods are constrained
to exceptional cases or the fortuitous eclipse geometry of binaries, and are complicated
by intrinsic variability (flares, new emerging flux) or the superposition of X-rays
from two objects. On the other hand, X-ray spectroscopic density measurements have opened a 
new  avenue toward assessing coronal structure.

He-like triplets show ratios between the ``forbidden'' and the ``intercombination'' 
line fluxes that are sensitive to the electron density $n_e$, the forbidden line being suppressed in high-density environments 
\cite{gabriel69,porquet01}. Under coronal temperature and density conditions,  transitions of 
C\,{\sc v}, N\,{\sc vi}, O\,{\sc vii}, Ne\,{\sc ix}, Mg\,{\sc xi}, and Si\,{\sc xiii} are
diagnostically useful, with some caveats. First, the density-sensitive range for the latter two
exceeds $n_{\rm e} \approx 10^{12}$~cm$^{-3}$, values that are probably uncommon to most solar-like stellar coronae.
Second, the densities to which He-like transitions are sensitive increases with the formation temperature
of the ion.
Low densities cannot be explicitly measured in very hot plasmas. And third, the maximum formation temperatures
of all ions mentioned above are in the range of 1--10~MK, while active stars  
reveal dominant electron temperatures up to several tens of MK (e.g., \cite{audard04}). Several Fe lines
are also sensitive to  $n_e$ (e.g., prominent lines of Fe\,{\sc  xvii} and  Fe\,{\sc xxi}; 
\cite{mewe85,brickhouse95}) 
although blends and inaccuracies in the atomic physics parameters make their application challenging. Fe lines
were already accessible to EUV spectroscopy performed by the EUVE satellite \cite{bowyer00}.

Triplet flux ratios measured in low-activity stars point to densities common to the solar corona, 
i.e., $n_e \approx 10^9 - 10^{10}$~cm$^{-3}$ (e.g., \cite{brinkman00,canizares00,mewe01,ness01,raassen02,raassen03}) at 
temperatures of a few MK. Magnetically active main-sequence stars show trends toward higher $n_e$, typically
of a few times  $10^{10}$~cm$^{-3}$, in the cool plasma as 
diagnosed for example by O\,{\sc vii} \cite{guedel01b,ness02a,raassen03b} while 
several subgiant active binaries indicate $n_e$ below the low-density limit 
of O~VII \cite{audard01,huenemoerder03,ness04}.

\begin{figure}
\vskip -0.5truecm
\psfig{file=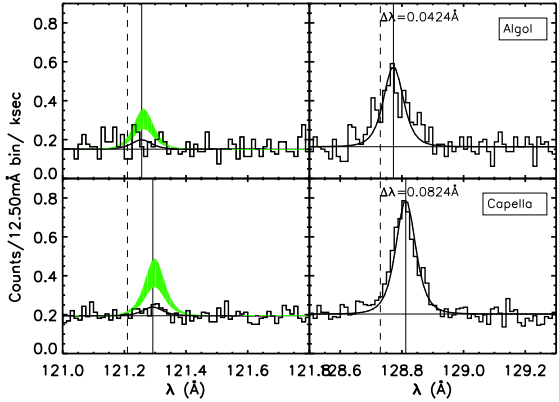,width=8.2cm}
\vskip -0.8truecm
\caption{Fe\,{\sc xxi} lines of Algol and Capella. The shaded areas indicate the expected 
    line shapes for a high-density plasma as reported previously (from \cite{ness04}). 
}\label{fig:ness04_f5a}
\vskip -0.7truecm
\end{figure}

A rather surprising - and to the present day controversial - result have been indications for {\it
very high} $n_e$ ($10^{12} - 10^{13}$~cm$^{-3}$) derived from ions forming in {\it hot} plasma 
(e.g., Mg\,{\sc xi}, Si\,{\sc xiii}, and Fe\,{\sc xxi}; see, e.g., \cite{mewe01,audard01b,argiroffi03}, 
and the work referring to EUVE summarized in ref. \cite{bowyer00}). The controversies are the following: 
\medskip

$\bullet$ Contradicting measurements using
different instruments; lines that should appear in high-density environments, and some of which were indeed 
reported from EUVE spectroscopy, are not present in high-resolution spectra obtained by
{\it Chandra}, indicating much lower $n_e$ (\cite{ayres01a,phillips01,ness04}, see 
Fig.~\ref{fig:ness04_f5a}). 

$\bullet$ Most high-density measurements 
straddle the low-density limit of the respective ion. 
This circumstance makes the measurements extremely vulnerable to  
inaccuracies in the atomic physics tabulations, and to unrecognized blends in some of
the lines. Slight modifications then have a dramatic effect on the implied $n_e$, as can be nicely seen in the 
analysis presented by ref. \cite{phillips01}. This work concluded, from a detailed study of 
the {\it Chandra} HETG spectrum of  Capella, that $n_e$ measured by the most 
reliable Fe\,{\sc xxi} line ratio, $f(\lambda 102.22)/f(\lambda 128.74)$, is compatible with the low-density 
limit of this diagnostic (i.e., $n_e < 10^{12}$~cm$^{-3}$). The similarity 
among the high density values just slightly above the low-density limits for a {\it wide variety of stars} 
points to a problem in the  atomic physics tabulations, while the true densities remain below the low-density limit 
\cite{ness04}. 

$\bullet$ Some $n_e$ measurements tend to be systematically different between ionization stages that 
have similar formation temperatures. In a study of the active binary $\sigma^2$ CrB, for example, densities 
derived from He-like triplets,  
Fe\,{\sc xxi} and Fe\,{\sc xxii}  reaches up to a few times $10^{12}$~cm$^{-3}$ whereas 
Mg\,{\sc xi} suggests $n_e < 10^{11}$~cm$^{-3}$ \cite{osten03}.

$\bullet$ Blends, perhaps from as of yet unrecognized lines, may bias the measured line-flux ratios 
systematically. The resolution of this issue requires very high spectral resolution. Detailed studies 
for the Ne\,{\sc ix} triplet and for the Mg\,{\sc xi} and Si\,{\sc xiii}
triplets  are available \cite{ness03b,testa04a}.
\medskip

\noindent If the density trend described above is real, then coronal loop pressures should vary by 3--4 
orders of magnitude. This obviously requires different magnetic loop systems for the different pressure 
regimes, with a tendency that hotter plasma occupies progressively smaller volumes \cite{osten03,argiroffi03,maggio04}.

\begin{figure}
\vskip -0.5truecm
\psfig{file=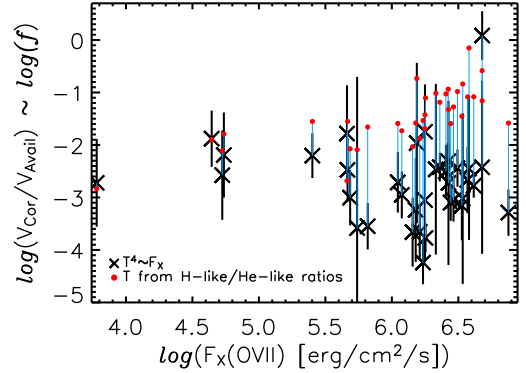,width=8cm}
\vskip -0.8truecm
\caption{Filling factors as a function of the O\,{\sc vii} flux, derived from densities measured with 
O\,{\sc vii} line-flux ratios, for a sample of nearby coronal stars. Different symbols refer to different 
scale height assumptions (from \cite{ness04}).
}\label{fig:ness04_f13a}
\vskip -0.7truecm
\end{figure}

The most comprehensive surveys of stellar coronal $n_e$ measurements are those given by refs. \cite{ness04} 
and \cite{testa04a}. These studies concluded that the surface filling factor
(derived from the emission measure, the measured $n_e$, and a realistic coronal scale height) of
magnetic loops containing {\it cool} X-ray emitting material increases from inactive to moderately active stars but then
``saturates'' at levels of about ten percent (Fig.~\ref{fig:ness04_f13a}). In the most active stars, hot coronal loops are added, with a sharply
increasing filling factor. This trend may be a consequence of increasing magnetic interaction in the corona. As one
moves from low to intermediate activity levels, more magnetic flux increases the number of active regions.
As the packing of active regions becomes denser, more frequent interactions between magnetic features induce 
more flare-like processes. The increased luminosity, the higher temperatures and the higher densities are then all a consequence of 
increased flaring and chromospheric evaporation in active regions \cite{guedel97,drake00}.

In general, the inversion problem from line flux ratios to density is 
degenerate, however. The simplest solution usually quoted, namely a plasma at constant density, is the 
least plausible one.
Realistic coronae reveal a distribution of electron densities, with  infinitely many distributions 
resulting in the same line flux ratios. The ratio-derived densities are not even linear averages 
across the emitting volume because of the $n_e^2$ dependence of the line fluxes. Denser volumes 
contribute disproportionately more X-rays than low-density volumes. The flux ratios thus provide information 
on possible distributions of volume in density \cite{guedel04}. 

\section{Dynamics from X-ray spectroscopy}

\begin{figure}
\vskip -0.1truecm
\psfig{file=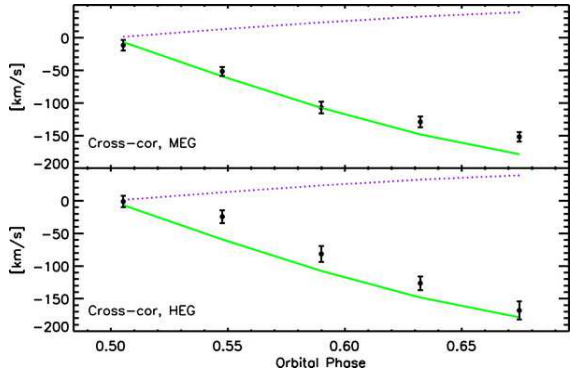,width=7.5cm}
\vskip -0.8truecm
\caption{Line-of-sight orbital velocity as a function of orbital phase, measured from shifts of
   X-ray lines in {\it Chandra} spectra of Algol. Solid lines indicate the expected velocities
   (from \cite{chung04}).
}\label{fig:chung04_f4}
\vskip -0.6truecm
\end{figure}

Doppler information from X-ray spectral lines may open up new ways of imaging 
 coronae of stars as they rotate, or as they orbit around the center of gravity in binaries.
First attempts are encouraging although the instrumental limitations are still severe.
Ref. \cite{ayres01b} reported Doppler shifts with amplitudes of $\approx 50$~km~s$^{-1}$
in X-ray lines of HR~1099. Amplitudes and phases clearly agreed with the line-of-sight orbital 
velocity of the subgiant K star, thus locating the bulk of the X-ray emitting plasma  on the latter
rather than in the intrabinary region.
Periodic line {\it broadening} in YY Gem, on the other hand, suggests that both components are 
similarly X-ray luminous \cite{guedel01a} which is expected because the two M dwarf components show
nearly identical fundamental properties. Ref. \cite{huenemoerder03}  found Doppler motions in
AR Lac to be compatible with coronae on both companions if the plasma is close to the photospheric 
level. For the  contact binary 44i Boo,  periodic line shifts corresponding to a total net velocity 
change over the full orbit of 180 km~s$^{-1}$ were found \cite{brickhouse01}. From the 
amplitudes and the phase of the rotational modulation \cite{brickhouse98}, 
the authors concluded that two dominant X-ray sources, one very compact and one extended, were present, 
both being located close to one of the stellar poles of the larger companion. 

\begin{figure}
\vskip -0.1truecm
\psfig{file=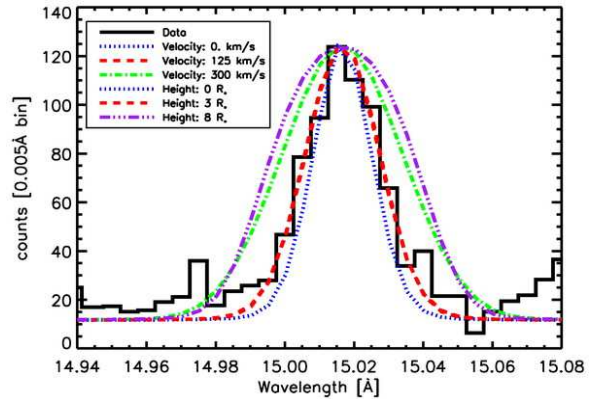,width=7.8cm}
\vskip -0.8truecm
\caption{Profile of the Fe\,{\sc xvii} 15.01~\AA\ line and various theoretical profiles, indicating significant
line broadening due to stellar rotation (from \cite{chung04}). 
}\label{fig:chung04_f6}
\vskip -0.6truecm
\end{figure}

A comprehensive study involving line shifts and broadening has been presented for the Algol binary 
\cite{chung04}. Periodic line shifts corresponding to a quadrature velocity of 150~km~s$^{-1}$
clearly prove that the X-rays are related to the K subgiant (Fig.~\ref{fig:chung04_f4}). However, 
the amplitude of the shifts
indicates that the source is slightly  displaced toward the 
B star, which may be the result of tidal distortions of the K star. 
Excess line broadening (above thermal and rotational broadening, Fig.~\ref{fig:chung04_f6}) 
can be ascribed to a radially extended corona,
with a coronal scale height of at least one stellar radius, which is consistent with expected 
scale heights of hot coronal plasma on this star.

\section{Abundances: I/FIP and solar models}

It is quite well established that the solar corona and the solar wind show elemental compositions 
at variance with the composition of the solar photosphere.
In the solar corona, elements with a  First Ionization Potential (FIP) above $\approx 10$~eV
(e.g., C, N, O, Ne, Ar) show photospheric abundance ratios with respect to hydrogen, while elements with a 
smaller FIP (e.g., Si, Mg, Ca, Fe) are overabundant by a factor of a few (``FIP Effect'', \cite{meyer85,feldman92}).  
Current thinking is that a fractionation process, probably 
 involving electric and/or magnetic fields, waves, or pressure gradients, occurs at chromospheric levels where 
 low-FIP elements are predominantly ionized and high-FIP elements are predominantly neutral (e.g., \cite{henoux95,laming04}).
Ions and neutrals can be affected differently by electric and magnetic fields.

\begin{figure}[t!]
\vskip -0.truecm
\psfig{file=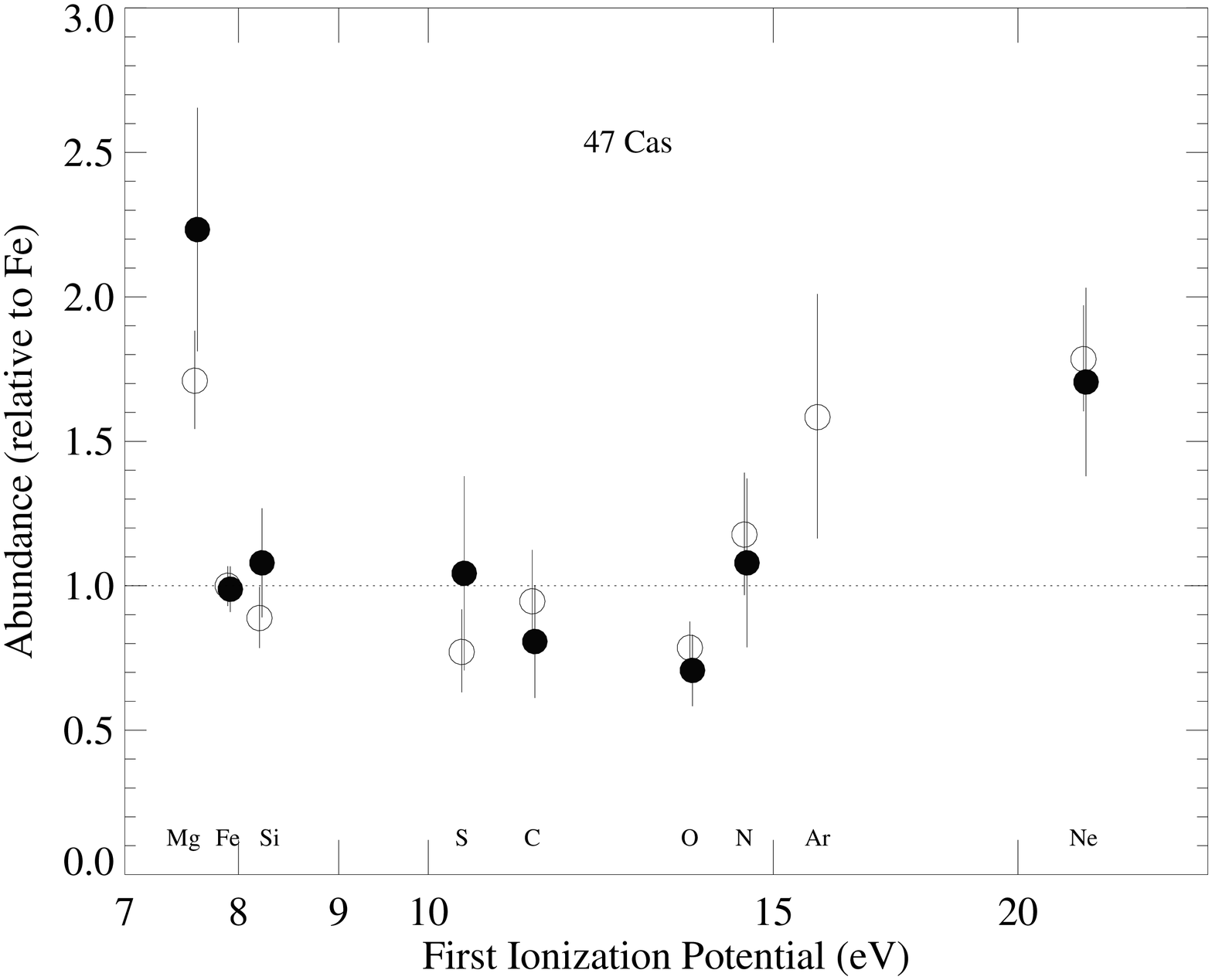,width=7.5cm}
\vskip 0.2truecm
\psfig{file=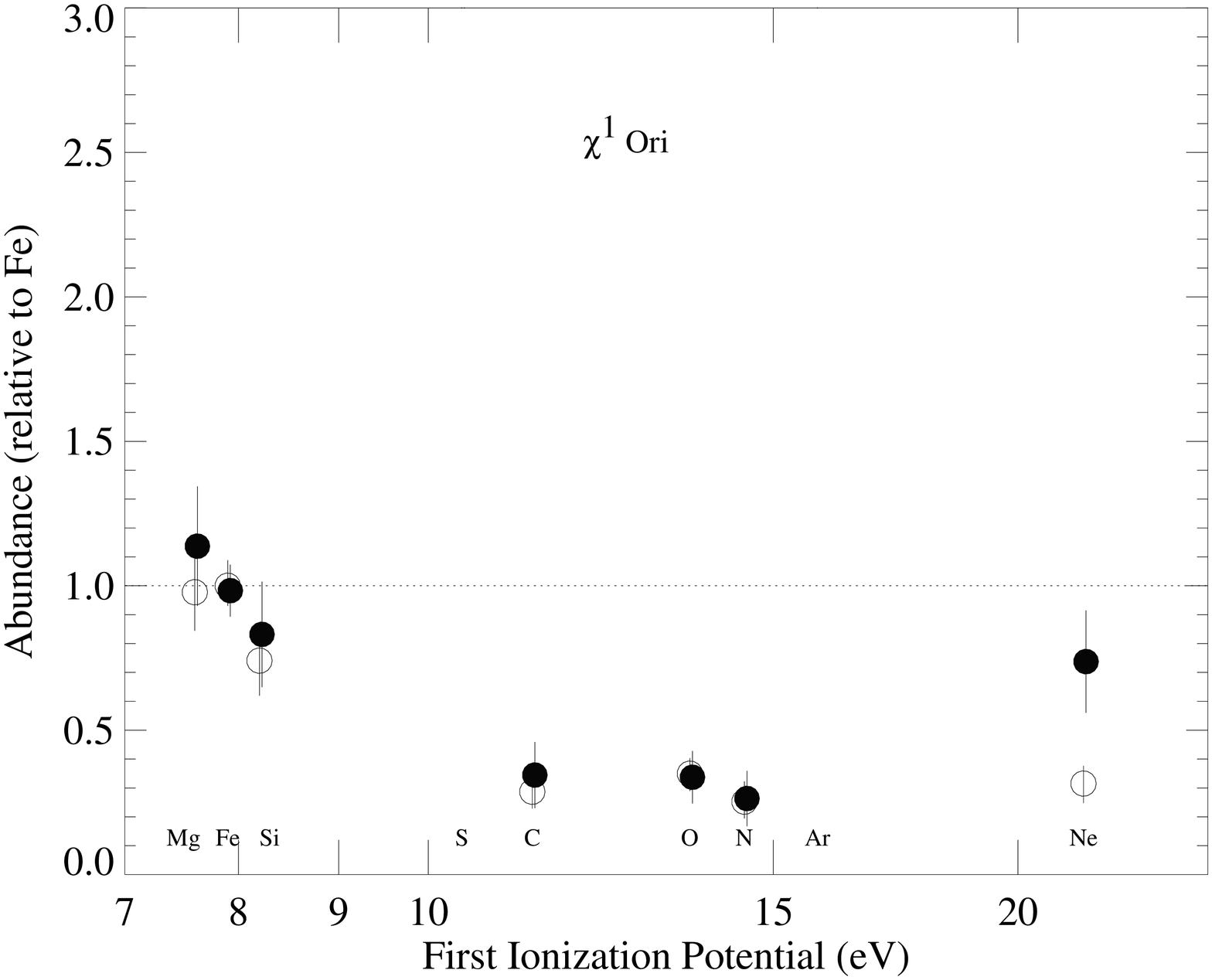,width=7.5cm}
\vskip -0.8truecm
\caption{Coronal  abundances as a function of FIP for the active solar analog 47 Cas B (top; with a 
mild IFIP trend) and for the more evolved $\chi^1$ Ori (bottom, with a FIP effect; from \cite{telleschi05}).
}\label{fig:guedel04_f36}
\vskip -0.8truecm
\end{figure}

First recognition of coronal abundance anomalies date back to the earliest stellar spectral X-ray observations (see summary 
in \cite{guedel04}). Clarification of trends with FIP, however, required 
high-resolution   grating spectroscopy. Early observations with  
{\it XMM-Newton} uncovered a new, systematic FIP-related
bias in magnetically active stars:  {\it low-FIP abundances are 
systematically depleted with respect to high-FIP elements} (\cite{brinkman01,guedel01b,audard01}; 
Fig.~\ref{fig:guedel04_f36}), a trend known as the ``inverse FIP effect'' (IFIP).
As a consequence of this anomaly, the ratio between the abundances of Ne (highest FIP) and Fe (low FIP)
is unusually large, up to about  10, compared to the solar photospheric ratio. These trends 
have been widely confirmed with the various gratings available on 
{\it XMM-Newton} or {\it Chandra},  for many active stars (e.g., \cite{drake01,huenemoerder01}). 

Systematic investigations have shown that the abundance pattern gradually changes from FIP to IFIP
with decreasing activity level (or increasing main-sequence age; \cite{audard03,telleschi05}, 
Fig.~\ref{fig:guedel04_f36}). In  nearby solar-analog stars, this effect does {\it not} mirror an abundance trend 
in the photosphere, which usually shows a composition in agreement with the solar photosphere \cite{telleschi05}. 

\begin{figure}[t!]
\vskip -0.4truecm
\psfig{file=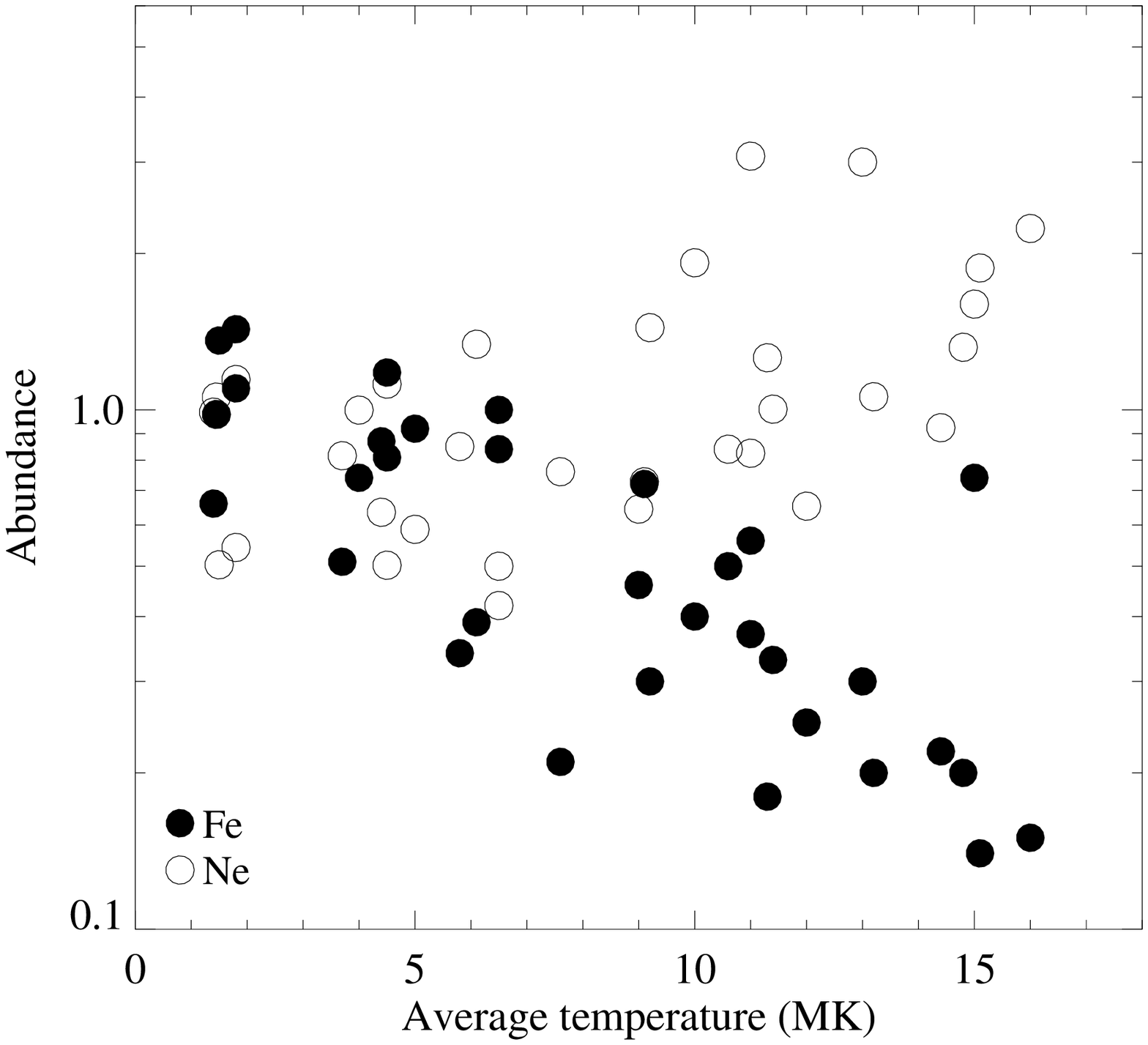,width=7.7cm}
\vskip -0.4truecm
\psfig{file=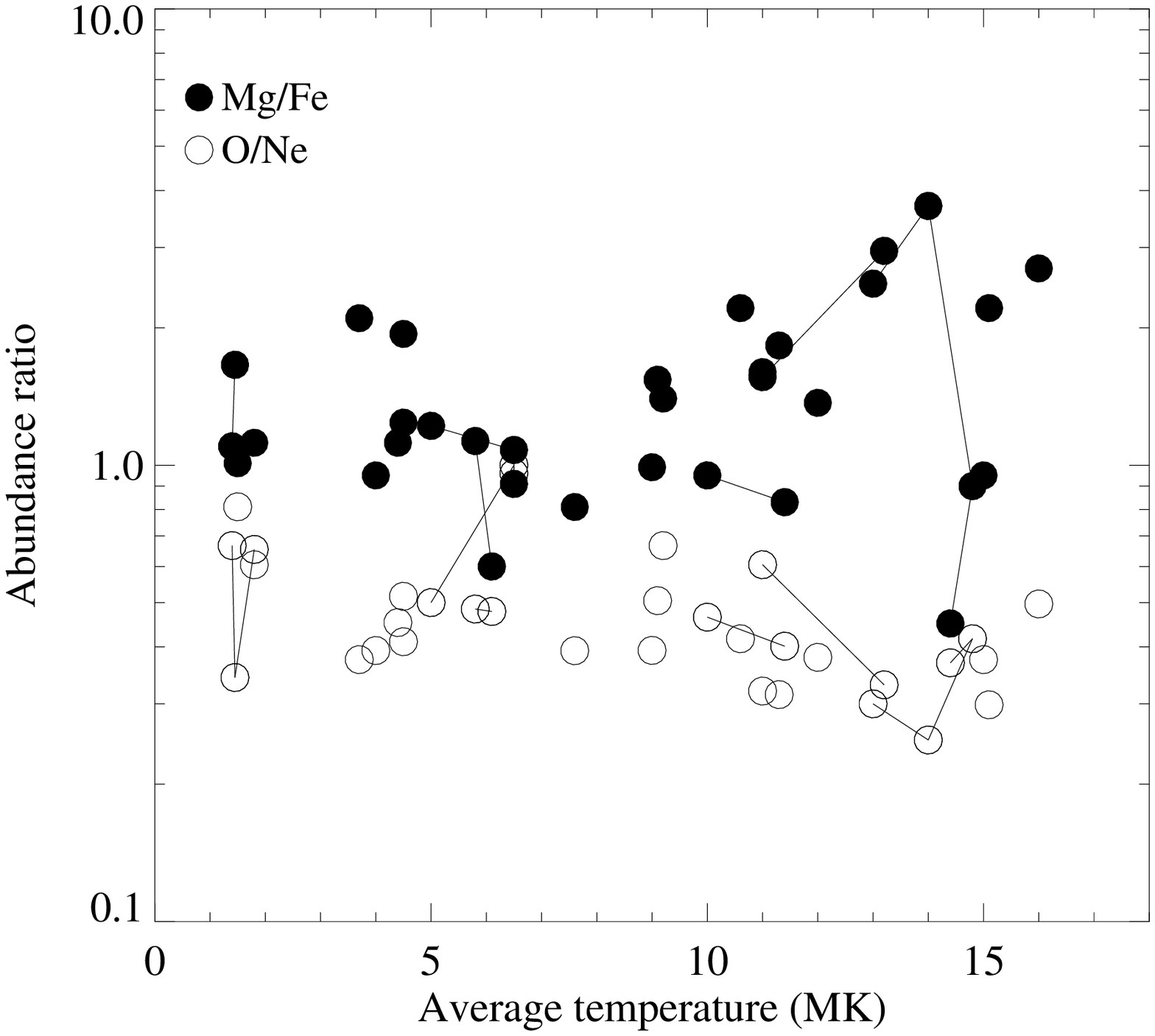,width=7.7cm}
\vskip -1.truecm
\caption{Coronal abundances of Fe and Ne (top), and ratios of Mg/Fe and O/Ne for various stars, shown as 
a function of the average coronal temperature (from \cite{guedel04}).
}\label{fig:guedel04_f37}
\vskip -0.6truecm
\end{figure}

Trends are summarized in Fig.~\ref{fig:guedel04_f37}  for the low-FIP elements Fe and Mg and the high-FIP elements Ne and O, as a function
of the average coronal temperature which serves as an activity indicator. Evidently, the absolute Fe abundance
decreases with increasing activity, while the Ne abundance slightly increases, leading to a largely
increasing Ne/Fe abundance ratio toward more active stars. The abundances reported here are quoted with respect
to the solar photospheric abundances reported in ref. \cite{anders89}, except for Fe that is taken from
ref. \cite{grevesse99}.

Fig.~\ref{fig:guedel04_f37} shows an anomaly for the O/Ne ratio which is found at values of 0.3-0.7 times the solar ratio,
apparently {\it regardless of the stellar activity level}. Because both O and Ne are high-FIP elements,
their abundance ratio {\it could} reflect the photospheric ratio. But then, the
Sun's composition would be anomalous. 

The tabulations of several solar element abundances  have recently
been revised, resulting in significant
discrepancies between solar models and helioseismological results (see ref. \cite{antia05} and references therein)
{\it unless} further solar abundances were also different. A Ne abundance higher by a factor of at least 2.5 than hitherto 
assumed  would be needed. Therefore,  ref. \cite{telleschi05} were first to  point out that the systematically non-solar 
O/Ne abundance ratio (by a similar factor) calls for a revision of the solar Ne abundance tabulation
which at the same time would solve the solar helioseismology problem. This was subsequently further 
elaborated by ref. \cite{drake05a} who suggested a factor of 2.7 upward revision of the adopted solar Ne abundance.

Two directions should be taken to verify this suggestion. First, further abundances of low-activity solar-analog
stars should be derived. So far, agreement with solar values have been reported for $\alpha$ Cen A \cite{raassen03},
Procyon \cite{raassen02}, and for $\beta$ Com, the latter with large error bars, however \cite{telleschi05}. There seems to
be a trend toward higher O/Ne ratios in Fig.~\ref{fig:guedel04_f37} indeed. New results for $\alpha$ Cen have been reported by
J. Schmitt (these proceedings). Second, verification of the solar Ne/O ratio is needed. Although Ne cannot be measured
directly in the photosphere, the Ne/O ratio can be derived from coronal measurements the same way as done for 
stellar observations. Recent analysis of solar active region X-ray spectra and of EUV spectra from
transition-region levels \cite{schmelz05,young05} both report the standard Ne/O abundance ratios,
rejecting an upward correction of Ne. The issue should therefore be considered to be open for the time being. 
A FIP-related enrichment of Ne in active corona remains  a viable possibility.

\section{Young stars: New phenomenology}

\subsection{Diskless and non-accreting systems}

As we move to younger stages, the {\it stellar environment}
becomes a key ingredient, made up of circumstellar  disks, gas flows from the disk
to the star perhaps along extended magnetospheric field lines, massive envelopes, and high-velocity jets
accompanied by massive, much slower molecular outflows. In recent years, surprising
and unexpected X-ray phenomenology related to these features has been uncovered.

\begin{figure}
\vskip -0.2truecm
\psfig{file=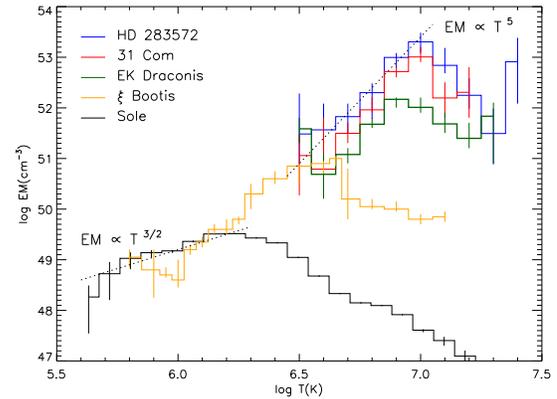,width=7.7cm}
\vskip -0.8truecm
\caption{Emission measure distributions of the WTTS HD~283572, the ZAMS star EK Dra, the giant 31 Com, the
G star $\xi$ Boo, and the Sun. Note  the steep slopes above $\log T \approx 6.5$ (from \cite{scelsi05}).
}\label{fig:scelsi05_f11}
\vskip -0.7truecm
\end{figure}

\begin{figure}
\vskip -0.4truecm
\psfig{file=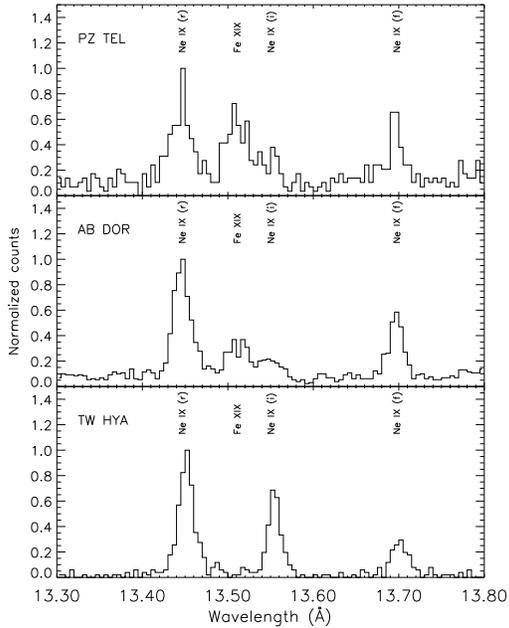,width=7cm}
\vskip -0.8truecm
\caption{Ne\,{\sc ix} triplets of PZ Tel, the ZAMS star AB Dor, and the CTTS TW Hya ({\it Chandra}  MEG spectra,
from top to bottom; from \cite{argiroffi04}).
}\label{fig:argiroffi04_8}
\vskip -0.7truecm
\end{figure}

\begin{figure}
\vskip -0.3truecm
\psfig{file=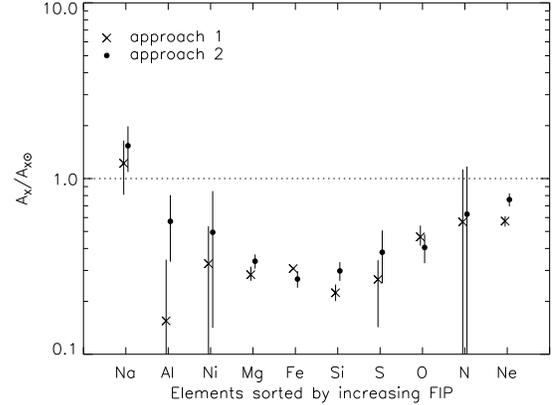,width=7.9cm}
\vskip -0.8truecm
\caption{Element abundances of PZ Tel as a function of FIP (from \cite{argiroffi04}).
}\label{fig:argiroffi04_4}
\vskip -0.7truecm
\end{figure}

Studies of stars that have dissipated their thick accretion disks and are evolving toward the main sequence
have almost uniformly revealed X-ray properties  in line with findings in young
main-sequence stars: very hot temperatures evident in broad emission-measure distributions (Fig.~\ref{fig:scelsi05_f11}), 
overall low densities (Fig.~\ref{fig:argiroffi04_8}), sub-solar abundances showing an IFIP effect 
(Fig.~\ref{fig:argiroffi04_4}), and flares.  
This holds true for rather evolved pre-main sequence objects such as the  rapidly rotating   
K0~V star PZ Tel  ($P = 0.94$~d, age 20~Myr; \cite{argiroffi04}, Fig.~\ref{fig:argiroffi04_8}, \ref{fig:argiroffi04_4}), 
but also for the much younger weak-lined T Tau stars (WTTS, \cite{scelsi05,kastner04}; Fig.~\ref{fig:scelsi05_f11}).

Two systems in the TW Hya Association (TWA), HD 98800 \cite{kastner04} and TWA~5 \cite{argiroffi05}, follow this trend
although both reveal unusually high Ne/Fe abundance ratios (up to 10). Although such ratios have been found in 
evolved binaries, they seem to be unusual  among WTTS.

\subsection{Classical T Tau stars: new physics?}

The emerging picture for non-accreting stars is that of a coronal origin of the X-rays in  analogy
to more evolved main-sequence stars. Does this hold for accreting classical T Tau (CTTS) stars?

\begin{figure}
\vskip -0.2truecm
\psfig{file=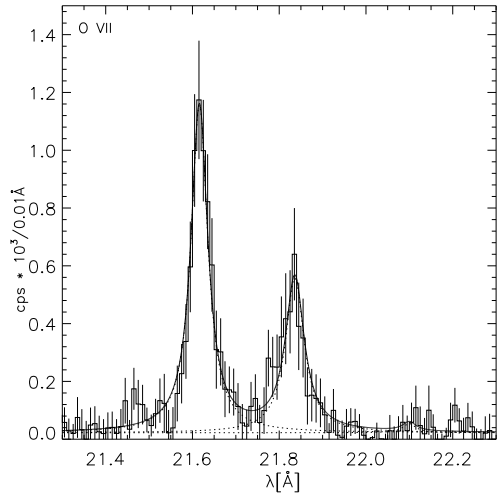,width=7.2cm}
\vskip -1.1truecm
\caption{O\, {\sc vii} triplet of TW Hya (from \cite{stelzer04}).
}\label{fig:stelzer04_f7b}
\vskip -0.7truecm
\end{figure}

The first CTTS studied in spectroscopic detail in X-rays was TW Hya in the TWA.  The high-resolution X-ray spectra from {\it Chandra} \cite{kastner02}
and {\it XMM-Newton} \cite{stelzer04} are full of surprises. First, and perhaps most notably, the emission line
pattern indicates a very cool plasma,  dominated by temperatures around 3~MK, with little hot plasma.
This is unusual for any type of T Tau stars. Second, the analysis of the Ne\,{\sc ix} and O\,{\sc vii} triplets reveals unusually
high densities of order $10^{12}-10^{13}$~cm$^{-3}$, the forbidden line of the O\,{\sc vii} line  essentially being absent 
(Fig.~\ref{fig:stelzer04_f7b}).
And third, TW Hya shows an unusually high Ne/Fe abundance ratio of about 10 relative to the solar photospheric ratio, 
and the N/O and N/Fe ratios are enhanced by a factor of 3.

The model proposed by ref. \cite{kastner02} and further elaborated in \cite{stelzer04} relies on magnetically
guided accretion streams from the disk to the star. Material falling toward the star reaches near-free fall
velocities, $v_{\rm ff} = (2GM/R)^{1/2}$, which is  a few hundred km~s$^{-1}$ for CTTS. Strong shocks 
at the photospheric impact site  heat material to
\begin{equation}
T = {3\mu m_{\rm H} v_{\rm ff}^2\over 16 k}
\end{equation}
where $k$ is the Boltzmann constant, $\mu$ is the mean molecular weight of the infalling plasma, and $m_{\rm H}$
is the mass of the hydrogen atom. The density developing in the shock is four times the pre-shock density.
The latter can be estimated if the global mass accretion rate, $\dot{M}$, and the accretion surface filling factor, $f$, are 
known, as follows:
\begin{equation}\label{acc}
n_{e, {\rm pre}} = {\dot{M} \over 4\pi R^2 f v_{\rm ff}\mu m_{\rm H}}
\end{equation}
where typical filling factors measured for T Tau stars are 0.1-10\% \cite{calvet98}. 
Shock temperatures of a few MK and electron densities $\gg 10^{10}$~cm$^{-3}$ are thus indeed expected.

The anomalous Ne/Fe (and N/Fe, O/Ne \cite{stelzer04,drake05b}) abundance ratios have been suggested to reflect depletion 
of Fe and O in the accretion disk where almost all elements condense into grains except for N \cite{savage96,charnley97} 
and Ne \cite{frisch03} that remain in the gas phase which is accreted onto the star.

\subsection{Cool plasma, Ne enhancements, high densities, or what?}

Why would a relatively evolved CTTS
with a rather moderate mass accretion rate show X-rays that are largely dominated by accretion-induced heating,
whereas other T Tau stars, either of equal age or younger examples in star-forming regions, show X-rays originating
predominantly from very hot (10-40~MK) coronal plasma (e.g., \cite{preibisch05})?  

To address this question further, two routes should be taken: i) a statistically meaningful sample of high-resolution 
spectra from T Tau stars should be collected, and ii) the anomalies claimed to relate to accretion should be critically
compared to stars with other properties.

Obtaining X-ray spectra from CTTS is challenging since only a handful are bright enough and not subject
to strong photoelectric absorption so as to allow access to cool material and the relevant He-like triplets.
BP Tau has been analyzed by ref. \cite{schmitt05}. While there is evidence for a suppressed forbidden line in O\,{\sc vii}
indicating electron densities of order $\log n_e = 11.5$ (lower than in TW Hya), the overall
plasma is dominated by very high temperatures. CR Cha has been reported to possibly  show a low forbidden line flux
in O\,{\sc vii} although the low signal-to-noise ratio makes this claim tentative \cite{robrade06}.

\begin{figure}
\vskip -0.0truecm
\psfig{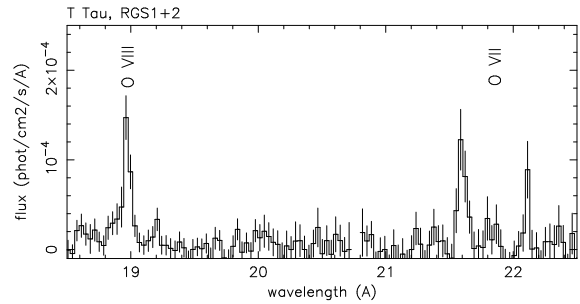}
\vskip -0.9truecm
\caption{{\it XMM-Newton} RGS spectrum of T Tau, showing the the O\,{\sc viii} Ly$\alpha$ and 
O\,{\sc vii} region.
}\label{fig:guedel_ttau}
\vskip -0.8truecm
\end{figure}

\begin{figure*}
\vskip -0.2truecm
\psfig{file=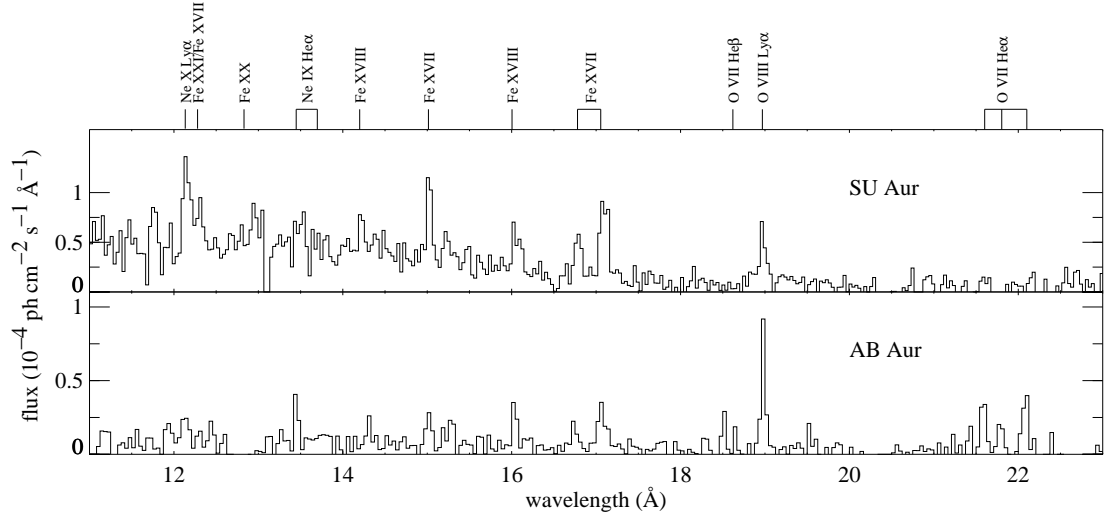,width=15cm}
\vskip -1.2truecm
\caption{RGS spectra of the CTTS SU Aur (top) and the Herbig star AB Aur (bottom; from \cite{telleschi06a}).
}\label{fig:telleschi_abaur}
\vskip -0.2truecm
\end{figure*}

Further grating spectra have been obtained as part of the {\it XMM-Newton Extended Survey of the Taurus Molecular Cloud}
(XEST, \cite{guedel06}). The example of T Tau is shown in Fig.~\ref{fig:guedel_ttau}. The ratio between its flux in 
the O\,{\sc vii} line triplet and the flux in the  O\,{\sc viii} Ly$\alpha$ line, O\,{\sc vii}/O\,{\sc viii},  is unusually
high even though the bulk of the emitting plasma is very hot. Such oxygen flux ratios are 
otherwise seen in rather evolved main-sequence stars, but not in active near-ZAMS objects \cite{telleschi05}.
The Herbig Ae star AB Aur, prototype of its class, reveals an RGS spectrum that is entirely dominated by cool
plasma, with a very prominent O\,{\sc vii} line triplet but faint Fe lines, the latter dominated by the lowest species
detectable in this range, namely Fe\,{\sc xvii} (Fig.~\ref{fig:telleschi_abaur}). Although winds are present in 
AB Aur, this object is thought to accrete matter at a similar rate as T Tau stars (see  \cite{telleschi06a}).

\begin{figure}[h!]
\psfig{file=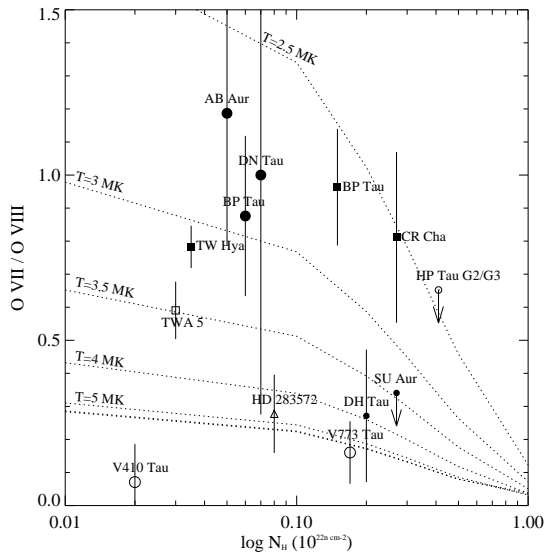,width=7.8cm}
\vskip -1truecm
\caption{Flux ratio O\,{\sc vii}/O\,{\sc viii} for CTTS (filled symbols) and WTTS 
        (open; from \cite{telleschi06b}). DH Tau was strongly flaring during the observations.
}\label{fig:telleschi_temp}
\vskip -0.7truecm
\end{figure}

Regardless of the overall, dominant electron temperature, there is thus evidence for a {\it soft excess} in CTTS compared 
to WTTS, best expressed by the O\,{\sc vii}/O\,{\sc viii} flux ratio. While this ratio is mostly $< 0.5$ in bright WTTS, 
it is around unity in CTTS. These systematics are illustrated in Fig.~\ref{fig:telleschi_temp} where the 
flux ratio is plotted as a function of the hydrogen absorption column density, $N_H$,  
allowing to estimate the unabsorbed flux (or luminosity) ratio and to model the equivalent isothermal plasma that would 
produce the same flux ratio \cite{telleschi06b}. {\it Accreting stars show a systematically higher ratio, indicating 
a soft excess.}

As for electron densities, apart from TW Hya and BP Tau, no further cases with a clear-cut suppression of the 
forbidden line in the O\,{\sc viii} or
Ne\,{\sc ix} triplets have been reported. On the contrary, both T Tau and AB Aur reveal low densities (the 1$\sigma$ error bar
reaching to $\approx 4\times 10^{10}$~cm$^{-3}$ for AB Aur, see Fig.~\ref{fig:telleschi_abaur};
Fig.~\ref{fig:guedel_ttau}  for T Tau).

A study of the element abundances in the X-ray sources of WTTS and CTTS, derived from {\it XMM-Newton} RGS 
spectroscopy, supports the often-reported trends of increasing abundances with 
increasing FIP \cite{telleschi06b}. Several CTTS {\it and}  WTTS show large Ne/Fe ratios ($\approx 4$ or higher), much larger than 
in main-sequence active solar analogs \cite{telleschi05} but similar to RS CVn binaries \cite{audard03}. 
In contrast, the CTTS SU Aur reveals a low  Ne/Fe abundance ratio of order unity \cite{robrade06,telleschi06b}, 
similar to some other massive CTTS \cite{telleschi06b}.
On the other hand, {\it high} Ne/Fe ratios are also found in non-accreting WTTS, for example in TWA 5 ($\approx 10$, as in TW Hya, \cite{argiroffi05}), 
HD~98800 ($\approx 5$  \cite{kastner04}), and in V773 Tau and V410 Tau \cite{telleschi06b}.

\begin{figure}
\psfig{file=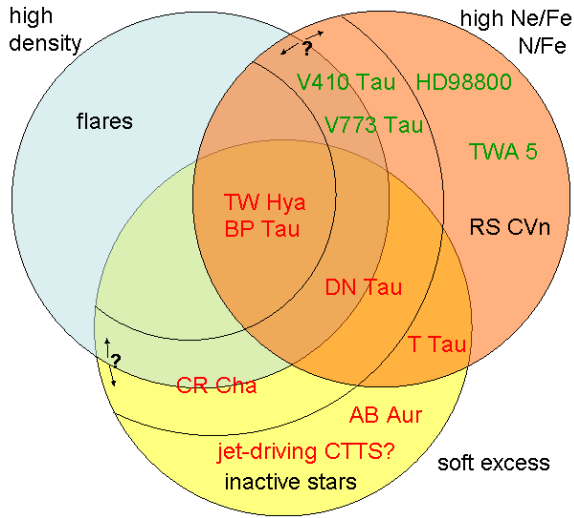,width=7.7cm}
\vskip -0.8truecm
\caption{T Tau stars sharing properties that have been claimed to be accretion indicators (high  $n_e$,
         soft plasma, Ne and N overabundances). CTTS are marked in red, WTTS in green. The annular region for
	 ``density'' contains objects for which density measurements are uncertain or not available.
}\label{fig:accretion}
\vskip -0.7truecm
\end{figure}

Fig.~\ref{fig:accretion} summarizes results from high-resolution spectroscopy of TTS with regard to the
three X-ray features claimed to be accretion indicators: high Ne/Fe abundance ratio, high electron densities, and soft excess. 
Only two stars,  TW Hya and BP Tau, reveal all three properties to some extent, although they are expressed considerably less in 
the latter. Ne/Fe abundance ratios do not seem to correlate with accretion, while {\it all} CTTS show a soft excess (as defined above).
(The existing grating spectra of SU Aur, HP Tau/G2, and HD 283572 contain insufficient information to place these stars onto
the diagram.)

What, then, are the possible origins of these features? Accretion-shock induced X-rays \cite{kastner02} remain a possibility
although the mechanism of X-ray escape remains problematic \cite{calvet98,drake05c}. The low O\,{\sc vii} derived
densities in  some of these  objects may relate to large surface accretion filling factors with consequent small mass 
flux densities and therefore small electron densities (see Eq.~\ref{acc}) although for stars like T Tau ($\dot{M} \approx 
4.5 \times 10^{-8}~M_{\odot}$~yr$^{-1}$, refs. in \cite{guedel06}), the accretion filling  factor would need to be around unity. In the Herbig star AB Aur, the 
O\,{\sc vii} forbidden line should be suppressed additionally by the photospheric radiation field, which is not observed. 
Its X-ray source cannot therefore be located close to the stellar surface \cite{telleschi06b}. It is also possible that
the observed O\,{\sc vii} triplets are dominated by coronal emission, outshining any contribution from high-density accretion-related
plasma. Obviously, this scenario is difficult to explain, in particular given the strong excess in the total O\,{\sc vii}
flux that is not usually observed in active coronal sources.

The most consistent trend in accreting stars is the {\it soft excess}. 
Explanations  include: i) accretion-shock heated material \cite{kastner02}; ii) elimination of the potentially hottest coronal loops by the accretion streams 
\cite{audard05}. iii) Additional
mass loading of coronal loops from the accretion streams, either suppressing the heating efficiency of these loops
\cite{preibisch05} or enhancing the cooling such that a soft excess builds up.

Another class of young soft-excess emitters have recently been identified in jet-driving CTTS. Their low-resolution 
spectra reveal a strong, little absorbed soft 
component from plasma at a few MK, and a strongly absorbed, flaring coronal component. Here, the accretion scenario is 
unlikely given the expected (and observed) strong absorption of X-rays from low-lying plasma. Rather, shocks
formed in the jets and outflows at some distance from the stars may be the origin of the X-rays \cite{guedel05,guedel06b}.

\section{Summary and Conclusions}

High-resolution X-ray spectroscopy has provided novel access to stellar coronae by resolving
emission lines from which the composition and structure of magnetically confined coronal
plasmas has been derived. There have also been little anticipated developments such as
X-rays related to the {\it stellar environment}, e.g. accretion disks, accretion flows, and
jets. An encouraging array of new emission models and of implications for stellar evolution
have been described in the literature. Nevertheless, limitations have rapidly been reached 
with the presently available grating spectrometers. Only few T Tau stars are accessible
to spectroscopic investigations, and some crucial classes of stars (protostars, brown dwarfs)
are out of reach of these gratings. The next generation of X-ray satellites should improve on
two parameters: wavelength resolution and sensitivity. A resolving power of a few 1000 will
systematically access orbit and rotational velocities of many stars and binaries. Flare
plasma flow velocities of a few 100~km~s$^{-1}$ can be systematically studied. Further, the problem
with unrecognized blends would be dramatically reduced. Sensitivity will be required to
study objects in star forming regions, in nearby open clusters, but also to access close
but faint objects such as evolved main-sequence stars or brown dwarfs.


\normalsize

\section*{ACKNOWLEDGEMENTS}
 
I thank C. Argiroffi, J. Drake, J.-U. Ness, L. Scelsi, B. Stelzer, and A. Telleschi for providing
illustrations shown in this article.                                           

\small
\begin{thebibliography}{9}
\bibitem{ottmann94}Ottmann, R. 1994, A\&A, 286, L27.
\bibitem{guedel95}G\"udel, M., Schmitt, J.~H.~M.~M., Benz, A.~O., \& Elias, N.~M. II. 1995, A\&A, 301, 201.
\bibitem{kuerster97}K\"urster, M., Schmitt, J.~H.~M.~M., Cutispoto, G., \& Dennerl, K. 1997, A\&A, 320, 831.    
\bibitem{audard01}Audard, M., G\"udel, M., \& Mewe, R. 2001, A\&A, 365, L318.
\bibitem{marino03}Marino, A., Micela, G., Peres, G., \& Sciortino, S. 2003, A\&A, 407, L63.
\bibitem{gabriel69}Gabriel, A.~H., \&  Jordan, C. 1969, MNRAS, 145, 241.
\bibitem{porquet01}Porquet, D., Mewe, R., Dubau, J., {\it et al.} 2001, A\&A, 376, 1113.
\bibitem{audard04}Audard, M., Telleschi, A., G\"udel, M., {\it et al.} 2004, ApJ, 617, 531.
\bibitem{mewe85}Mewe, R., Gronenschild, E.~H.~B.~M., \& van den Oord, G.~H.~J. 1985, A\&AS  62, 197.
\bibitem{brickhouse95}Brickhouse, N.~S., Raymond, J.~C., \& Smith, B.~W. 1995, ApJS, 97, 551.
\bibitem{bowyer00}Bowyer, S., Drake, J.~J., \& Vennes, S.  2000, ARA\&A, 38, 231.
\bibitem{brinkman00}Brinkman, A.~C., Gunsing, C.~J.~T., Kaastra, J.~S., {\it et al.} 2000, ApJ, 530, L111. 
\bibitem{canizares00}Canizares, C.~R., Huenemoerder, D.~P., Davis, D.~S., {\it et al.} 2000, ApJ, 539, L41.  
\bibitem{mewe01}Mewe, R., Raassen, A.~J.~J., Drake, J.~J., {\it et al.} 2001, A\&A, 368, 888.
\bibitem{ness01}Ness, J.-U., Mewe, R., Schmitt, J.~H.~M.~M., {\it et al.} 2001, A\&A, 367, 282.    
\bibitem{raassen02}Raassen, A.~J.~J., Mewe, R., Audard, M., {\it et al.} 2002, A\&A, 389, 228.
\bibitem{raassen03}Raassen, A.~J.~J., Ness, J.-U., Mewe, R., {\it et al.} 2003, A\&A, 400, 671.
\bibitem{guedel01b}G\"udel, M., Audard, M., Briggs, K., {\it et al.} 2001b, A\&A, 365, L336.
\bibitem{ness02a}Ness, J.-U., Schmitt, J.~H.~M.~M., Burwitz, V., {\it et al.} 2002a, A\&A, 394, 911. 
\bibitem{raassen03b}Raassen, A.~J.~J., Mewe, R., Audard, M., \& G\"udel, M. 2003b, A\&A, 411, 509.
\bibitem{huenemoerder03}Huenemoerder, D.~P., Canizares, C.~R., Drake, J.~J., \& Sanz-Forcada, J. 2003, ApJ, 595, 1131. 
\bibitem{ness04}Ness, J.-U., G\"udel, M., Schmitt, J.~H.~M.~M., {\it et al.} 2004, A\&A, 427, 667.  
\bibitem{audard01b}Audard, M., Behar, E., G\"udel, M., {\it et al.} 2001b, A\&A, 365, L329.
\bibitem{argiroffi03}Argiroffi, C., Maggio, A.,  \& Peres, G. 2003, A\&A,  404, 1033. 
\bibitem{ayres01a}Ayres, T.~R., Osten, R.~A., \& Brown, A. 2001a, ApJ, 562, L83.
\bibitem{phillips01}Phillips, K.~J.~H., Mathioudakis, M., Huenemoerder, D.~P., {\it et al.} 2001, MNRAS, 325, 1500.
\bibitem{osten03}Osten, R.~A., Ayres, T.~R., Brown, A., {\it et al.} 2003, ApJ, 582, 1073.
\bibitem{ness03b}Ness, J.-U., Brickhouse, N.~S., Drake, J.~J., \& Huenemoerder, D.~P. 2003b, ApJ, 598, 1277.  
\bibitem{testa04a}Testa, P., Drake J.~J., \& Peres, G. 2004a, ApJ, 617, 508.
\bibitem{maggio04}Maggio, A., Drake J.~J., Kashyap, V.~L., {\it et al.}  2004, ApJ, 613, 548.
\bibitem{guedel97}G\"udel, M., Guinan, E.~F., \&  Skinner, S.~L. 1997, ApJ, 483, 947.
\bibitem{drake00}Drake, J.~J., Peres, G., Orlando, S., {\it et al.} 2000, ApJ, 545, 1074. 
\bibitem{guedel04}G\"udel, M. 2004, A\&ARev, 12, 71.
\bibitem{ayres01b}Ayres, T.~R., Brown, A., Osten, R.~A., {\it et al.} 2001b, ApJ, 549, 554.   
\bibitem{guedel01a}G\"udel, M., Audard, M., Magee, H., {\it et al.}  2001a, A\&A, 365, L344.    
\bibitem{brickhouse01}Brickhouse,  N.~S., Dupree, A.~K., \& Young, P.~R. 2001, ApJ, 562, L75.
\bibitem{brickhouse98}Brickhouse, N.~S., \& Dupree, A.~K. 1998, ApJ, 502, 918.
\bibitem{chung04}Chung, S.~M., Drake, J.~J., Kashyap, V.~L., {\it et al.}  2004, ApJ, 606, 1184.
\bibitem{meyer85}Meyer, J.-P. 1985, ApJS, 57, 151.
\bibitem{feldman92}Feldman, U. 1992, Phys. Scripta, 46, 202.
\bibitem{henoux95}H\'enoux, J.-C. 1995, Adv. Space Res., 15, 23.
\bibitem{laming04}Laming, M. 2004, ApJ, 614, 1063.
\bibitem{brinkman01}Brinkman, A.~C., Behar, E., G\"udel, M., {\it et al.} 2001, A\&A, 365, L324.  
\bibitem{drake01}Drake, J.~J., Brickhouse, N.~S., Kashyap, V., {\it et al.}  2001, ApJ, 548, L81.
\bibitem{huenemoerder01}Huenemoerder, D.~P., Canizares, C.~R.,  \& Schulz, N.~S. 2001, ApJ, 559, 1135.
\bibitem{audard03}Audard, M., G\"udel, M., Sres, A., {\it et al.} 2003, A\&A, 398, 1137.
\bibitem{telleschi05}Telleschi, A., G\"udel, M.,  Briggs, K., {\it et al.} 2005, ApJ, 622, 653. 
\bibitem{anders89} Anders, E., \& Grevesse, N. 1989, Geochim. Cosmochim. Acta, 53, 197.
\bibitem{grevesse99}Grevesse, N.,  \& Sauval, A.~J. 1999, A\&A, 347, 348.  
\bibitem{antia05}Antia, H.~M., \& Basu, S. ApJ, 620, L129.
\bibitem{drake05a}Drake, J.~J., \& Testa, P. 2005, Nature, 436, 525.  
\bibitem{schmelz05}Schmelz, J.~T., Nasraoui, K., Roames, J.~K., {\it et al.} 2005, ApJ, 634, L197.
\bibitem{young05}Young, P.~R. 2005, A\&A, 444, L45.  
\bibitem{argiroffi04}Argiroffi, C., Drake, L.~J., Maggio, A., {\it et al.}  2004, ApJ,  609, 925. 
\bibitem{scelsi05}Scelsi, L., Maggio, A.,  Peres, G., \& Pallavicini, R. 2005, A\&A, 432, 671. 
\bibitem{kastner04}Kastner, J.~H., Huenemoerder, D.~P., Schulz, N.~S., {\it et al.} 2004, ApJ, 605, L49.  
\bibitem{argiroffi05}Argiroffi, C., Maggio, A., Peres, G., {\it et al.}  2005, A\&A,  439, 1149. 
\bibitem{kastner02}Kastner, J.~H., Huenemoerder, D.~P., Schulz, N.~S., {\it et al.} 2002, ApJ, 567, 434.
\bibitem{stelzer04}Stelzer, B., \& Schmitt, J.~H.~M.~M. 2004, A\&A, 418, 687.  
\bibitem{calvet98}Calvet, N.,  \& Gulbring, E. 1998, ApJ, 509, 802.
\bibitem{drake05b}Drake, J.~J., Testa, P., \& Hartmann, L. 2005, ApJ, 627, L149. 
\bibitem{savage96}Savage, B.~D., \& Sembach,  K.~R. 1996, ARA\&A, 34, 279.
\bibitem{charnley97}Charnley, S.~B. 1997, MNRAS, 291, 455.
\bibitem{frisch03}Frisch, P.~C., \& Slavin, J.~D. 2003, ApJ, 594, 844.
\bibitem{preibisch05}Preibisch, T., Kim, Y.-C., Favata, F., et al. 2005,  ApJS, 160, 401. 
\bibitem{schmitt05}Schmitt, J.~H.~M.~M., Robrade, J., Ness, J.-U., {\it et al.} 2005,  A\&A, 432, L35. 
\bibitem{robrade06}Robrade, J., \& Schmitt, J.~H.~M.~M. 2006,  A\&A, 449, 737.
\bibitem{guedel06}G\"udel, M., Briggs, K.~R., Arzner, K., {\it et al.}  2006,  A\&A, submitted.
\bibitem{telleschi06a}Telleschi, A., G\"udel, M., Briggs, K.~R., {\it et al.}  2006a,  A\&A, submitted. 
\bibitem{telleschi06b}Telleschi, A., G\"udel, M., Briggs, K.~R., {\it et al.}  2006b,  A\&A, submitted. 
\bibitem{drake05c}Drake, J.~J. 2005, in Cool Stars 14, eds. F. Favata et al. (Noordwijk: ESA), 519.
\bibitem{audard05}Audard, M., G\"udel, M., Skinner, S.~L., {\it  et al.}  et al. 2005, ApJ, 635, L81. 
\bibitem{guedel05}G\"udel, M., Skinner, S.~L., Briggs, K.~R.,  et al. 2005, ApJ, 626, L53.
\bibitem{guedel06b}G\"udel, M.,Telleschi, A., Audard, M., {\it et al.}  2006,  A\&A, submitted.

\end{thebibliography}
\end{document}